\documentclass[12pt]{iopart}

\newif\ifanonymized
\anonymizedfalse

\newif\ifshowupdates
\showupdatesfalse
\ifshowupdates
\newcommand{\update}[1]{\textcolor{red}{#1}}
\else
\newcommand{\update}[1]{#1}
\fi

\newif\ifincludeappendix
\includeappendixtrue

\expandafter\let\csname equation*\endcsname\relax
\expandafter\let\csname endequation*\endcsname\relax

\usepackage{amsmath}
\usepackage{amssymb}
\usepackage{mathtools}
\usepackage{csquotes}
\usepackage{graphicx}
\usepackage{dcolumn}
\usepackage{bm}
\usepackage{hyperref}
\usepackage{xcolor}
\usepackage{tabularx}
\usepackage{booktabs}
\usepackage{cite}

\usepackage{tikz}
\usepackage{pgfplots}
\usetikzlibrary {arrows.meta,bending,positioning,calc}

\newcommand{\avgx}[2]{{\mathbb E}_{#2} \left [  {#1} \right]}
\newcommand{\abs}[1]{\left|{#1}\right|}
\newcommand{\kl}[2]{D\!\left({#1}\,\|\,{#2}\right)}
\newcommand{\nlog}[1]{\ln{#1}}
\newcommand{\defeq}{\vcentcolon=}
\newcommand{\g}{\,|\,}
\newcommand{\gauss}{\mathcal{N}}
\newcommand{\ourmodel}{$NPT$-flow}

\begin{document}

\title{\update{Estimating} Gibbs free energies via isobaric-isothermal flows}

\ifanonymized
\author{Anonymous authors}
\else
\author{Peter Wirnsberger\footnote{Author to whom any correspondence should be addressed: \mailto{pewi@google.com}.}, Borja Ibarz, George Papamakarios}
\address{DeepMind, London, United Kingdom}
\fi

\begin{abstract}
We present a machine-learning model based on normalizing flows that is trained to sample from the isobaric-isothermal ensemble. In our approach, we approximate the joint distribution of a fully-flexible triclinic simulation box and particle coordinates to achieve a desired internal pressure. This novel extension of flow-based sampling to the isobaric-isothermal ensemble yields direct estimates of Gibbs free energies. We test our $NPT$-flow on monatomic water in the cubic and hexagonal ice phases and find excellent agreement of Gibbs free energies and other observables compared with established baselines.
\end{abstract}

\section{Introduction\label{sec:intro}}

Computer simulations have become an indispensable tool to refine our theoretical understanding of matter.
Molecular dynamics \update{(MD)} and Monte Carlo methods are the two bedrock techniques for molecular simulation~\cite{FrenkelSmit}. Recent advances in machine learning (ML), however, have enabled another paradigm of sampling from high-dimensional Boltzmann distributions that is based on powerful generative models, in particular normalizing flows~\cite{Rezende2015, Noe2019, Papamakarios2021, Tuckerman2019}.
An appealing property of normalizing flows is that they support drawing independent samples and provide exact likelihoods that are, for example, useful for efficient free energy estimation~\cite{Wirnsberger2020, Nicoli2020}. Irrespective of the sampling method, statistical mechanics provides us with the mathematical language to reason about the complex interactions within a probabilistic framework.

A key goal in equilibrium statistical mechanics is the efficient estimation of expectation values $\avgx{A(x)}{p(x)}$, where $A(x)$ is an arbitrary function of a typically high-dimensional vector $x$ and $p(x)$ is the probability density which depends on a set of constant external parameters that define the \textit{statistical ensemble}~\cite{FrenkelSmit}. Two popular ensembles are the \textit{canonical} ensemble, at fixed number of particles $N$, volume $V$ and temperature $T$, and the \textit{isobaric-isothermal} ensemble, at fixed $N$, $P$ and $T$, where $P$ is the pressure.
While most of the differences between ensembles disappear in the \enquote{thermodynamic limit}, where $N \to \infty$ and $N/V \to \rho$ with $\rho$ being the density, fluctuations and statistical averages can differ for some observables~\cite{FrenkelSmit}. It is therefore crucial to sample from the correct distribution that is compatible with the external set of parameters.
 
Despite the considerable interest in ML-based sampling of molecular systems, applications of normalizing flows have primarily focused on the canonical ensemble~\cite{Noe2019, Wirnsberger2020, Gabrie2022, Nicoli2021, Ahmad2022, Wirnsberger2022normalizing}, and do not support sampling from the isobaric-isothermal ensemble. Imposing a constant pressure is crucial, however, for studying important physical phenomena such as first-order phase transitions, across which the pressure is constant but the volume changes. It is further important for studying solid crystals to ensure that they can relax to their equilibrium shape~\cite{Parrinello1981, Vega2008}. Moreover, a normalizing flow capable of sampling from the isobaric-isothermal ensemble would provide us with a simple and direct route to estimating the Gibbs free energy, by utilizing it as a targeted free energy estimator~\cite{Jarzynski2002, Wirnsberger2020}.

In this work, we propose a normalizing flow model, called \ourmodel{}, that jointly generates particle coordinates and triclinic box shape parameters \update{keeping the number of particles $N$, the pressure $P$ and the temperature $T$ fixed. The model can therefore be} trained to approximate the target density of the isobaric-isothermal ensemble. The paper is structured as follows. First, we provide some background on flow-based sampling of the canonical ensemble in Section~\ref{sec:background} and introduce the $NPT$-flow for the isobaric-isothermal ensemble in Section~\ref{sec:method}. We then benchmark the model on ice in two different crystal phases by comparison with MD data in Section~\ref{sec:results}, and discuss limitations of and interesting extensions to our approach in Section~\ref{sec:discussion}.

\section{Background\label{sec:background}}

This section discusses how normalizing flows can be used for unbiased estimation of observables within the framework of statistical mechanics. The general idea is to train a normalizing flow to closely approximate the canonical ensemble using standard deep-learning tools, and then estimate expectations under the ensemble using the flow model as a proposal distribution in an importance-weighting scheme. Readers familiar with this background can skip to Section~\ref{sec:method}, where the extension of this methodology to the isobaric-isothermal ensemble is presented.

\subsection{Canonical ensemble and Helmholtz free energy}

We consider a system of $N$ atoms in $D$ dimensions, confined in a fixed simulation box (typically orthorhombic). Let $x_n = (x_{n,1}, \ldots, x_{n,D})$ be the position coordinates of the $n$-th particle, and $x = (x_1, \ldots, x_N)$ be the configuration of the system. We assume that the atoms are subject to a potential energy $U(x)$ and the system is in contact with a heat bath at fixed temperature $T$ with which it can exchange energy. A standard result in statistical mechanics is that the equilibrium distribution of such a system (known as the \textit{canonical} or $NVT$ ensemble) is given by the Boltzmann distribution:
\begin{equation}
p(x) = \frac{1}{Z_V}\exp[-\beta U(x)],
\label{eq:boltzmann_distribution}
\end{equation}
where $Z_V = \int \mathrm dx \exp[-\beta U(x)]$ is the canonical partition function and $\beta = 1/k_\mathrm{B} T$ is the inverse temperature ($k_\mathrm{B}$ is Boltzmann's constant). The canonical partition function is directly related to the \textit{Helmholtz free energy} $F$ via
\begin{equation}
F = -\beta^{-1}\nlog{Z_V}.
\end{equation}

In practice, one is interested in estimating (i) expectations under the Boltzmann distribution $\avgx{A(x)}{p(x)}$ of physical observables $A$, and (ii) the partition function $Z_V$ and hence the free energy $F$. These problems are computationally challenging due to the high dimensionality of $x$ and the fact that $p(x)$ can only be evaluated point-wise and up to a constant. In what follows, we focus on a method that takes advantage of recent advances in machine learning and deep generative modelling, specifically \textit{normalizing flows}.

\subsection{Normalizing flows}

Normalizing flows are deep generative models for parameterizing flexible probability distributions. A normalizing flow parameterizes a complex distribution $q_\theta(x)$ (with parameters $\theta$) as the pushforward of a (typically fixed) simple base distribution $q(u)$ via an invertible and differentiable function $f_\theta$ (typically parameterized with deep neural networks). In other words, a sample $x$ from $q_\theta(x)$ is drawn by first sampling $u$ from $q(u)$ and transforming it with $f_\theta$:
\begin{equation}
    x \sim q_\theta(x) \quad\Leftrightarrow\quad x = f_\theta(u), u\sim q(u).
    \label{eq:flow_sampling}
\end{equation}
Thanks to the invertibility and differentiability of $f_\theta$, the probability density of $x$ can be computed exactly with a change of variables:
\begin{equation}
    q_\theta(x) = q(u)\abs{\det \frac{\partial f_\theta}{\partial u}}^{-1},
\end{equation}
where $\frac{\partial f_\theta}{\partial u}$ is the Jacobian of $f_\theta$. There is a wide variety of implementations of invertible and differentiable functions $f_\theta$ using deep neural networks that are flexible enough to parameterize complex distributions $q_\theta(x)$ but whose Jacobian determinant $\det \frac{\partial f_\theta}{\partial u}$ remains efficient to compute, enabling exact density evaluation; for a thorough review, see \cite{Papamakarios2021, Kobyzev2021}.

\subsection{Training normalizing flows to approximate target distributions}

Given a target distribution that can be evaluated point-wise and up to a constant, such as the Boltzmann distribution in Eq.~\eqref{eq:boltzmann_distribution}, the objective is to train a normalizing-flow model to approximate it. 
Let the intractable target distribution take the general form $p(x) = \frac{1}{Z}\exp[g(x)]$ and let $q_\theta(x)$ be a flow model as defined in Eq.~\eqref{eq:flow_sampling}. We train the flow to approximate the target by minimizing a suitably chosen divergence between $p(x)$ and $q_\theta(x)$ with respect to $\theta$. A common choice is the reverse Kullback--Leibler divergence:
\begin{align}
    \label{eq:kl}
    \kl{q_\theta(x)}{p(x)} &= \avgx{\nlog{q_\theta(x)} - \nlog{p(x)}}{q_\theta(x)} \\ 
    &= \avgx{\nlog{q_\theta(x)} - g(x)}{q_\theta(x)} + \nlog{Z}   \\
    &= \avgx{\nlog{q(u)} - \nlog{\abs{\det \frac{\partial f_\theta}{\partial u}}} -  g(f_\theta(u))}{q(u)} + \nlog{Z},
\end{align}
where in the last step we reparameterized the expectation with respect to the base distribution $q(u)$. Using the final expression, the gradient of the KL divergence with respect to $\theta$ is
\begin{equation}
\nabla_\theta \kl{q_\theta(x)}{p(x)} = \avgx{ - \nabla_\theta\nlog{\abs{\det \frac{\partial f_\theta}{\partial u}}} -  \nabla_\theta g(f_\theta(u))}{q(u)}.
\end{equation}
Using samples from $q(u)$, one can readily obtain an unbiased estimate of the above gradient, which can be used to minimize the KL divergence with any stochastic gradient optimization method.

\subsection{Estimation of observables and free energy}

After we train a flow model $q_\theta(x)$ to approximate the target distribution $p(x)$, samples from $q_\theta(x)$ can be used in a Monte-Carlo fashion to estimate expectations $\avgx{A(x)}{p(x)}$. Since $q_\theta(x)$ does not exactly equal $p(x)$, direct Monte-Carlo estimates with samples from $q_\theta(x)$ will be biased; to avoid that, the probability density of samples can be used to remove the bias. Assuming $q_\theta(x) > 0$ whenever $p(x) > 0$, the simplest way to remove bias is with importance weighting: using i.i.d.\ samples $(x^{(1)}, \ldots, x^{(S)})$ from $q_\theta(x)$, we can form the asymptotically unbiased estimate
\begin{equation}
 \avgx{A(x)}{p(x)} \approx \frac{\sum_{s=1}^S w(x^{(s)})A(x^{(s)})}{\sum_{s=1}^S w(x^{(s)})}
 \quad\text{where}\quad   w(x) \defeq \frac{\exp[g(x)]}{q_\theta(x)}.
\end{equation}
Similarly, the normalizing constant $Z = \int \mathrm{d}x \exp[g(x)]$ can be unbiasedly estimated as
\begin{equation}
 Z \approx \frac{1}{S}\sum_{s=1}^S w(x^{(s)}).
 \label{eq:fep}
\end{equation}
\update{A disadvantage of importance-weighting is that it results in an estimator whose statistical performance is lower than that of a standard Monte-Carlo average of $S$ independent samples. To quantify this reduction in statistical performance, the effective sample size (ESS) is often used, estimated by~\cite{kong1992ess}}
\begin{equation}
    \update{\mathrm{ESS} \approx \frac{\left[\sum_{s=1}^S w(x^{(s)})\right]^2}{\sum_{s=1}^S \left[w(x^{(s)})\right]^2}.}
\end{equation}
\update{The ESS is a number between $1$ and $S$ (often reported as a percentage of $S$), and roughly measures how many independent samples would result in a Monte-Carlo estimator of the same statistical performance.}

In statistical mechanics, the estimation method in Eq.~\eqref{eq:fep} is widely known as \textit{free energy perturbation} (FEP)~\cite{Zwanzig1954}; when combined with a configuration space map it is referred to as targeted free energy perturbation (TFEP)~\cite{Jarzynski2002} and, in combination with a learned map or a flow model it is referred to as \textit{learned free energy perturbation} (LFEP)~\cite{Wirnsberger2020, Nicoli2020}.

In general, the simpler the estimation method, the closer the flow model must approximate the target for the statistical efficiency of the estimator to be within practical constraints. More sophisticated, but also computationally more expensive, multi-step estimation methods are available and can be utilized as learned estimators\update{~\cite{Wirnsberger2020, Arbel2021, Matthews2022, Rizzi2023, Midgley2023}} if higher accuracy is desirable.

\section{Method\label{sec:method}}

This section extends the application of normalizing flows from the canonical to the isobaric-isothermal ensemble with triclinic simulation boxes. The presentation is in two parts: Section~\ref{sec:isothermal_isobaric_ensemble} shows how to construct a target distribution for the isobaric-isothermal ensemble, and Section~\ref{sec:npt_flow} proposes a flow architecture suitable for this target.

\subsection{Isobaric-isothermal ensemble for atomistic systems\label{sec:isothermal_isobaric_ensemble}}

In the isobaric-isothermal ensemble, the external pressure $P$ is held fixed while the shape of the simulation box, and hence its volume $V$, fluctuates. Therefore, the shape parameters of the simulation box should be treated as additional degrees of freedom that co-vary with the configuration $x$. In what follows, we will assume that the shape of the simulation box is controlled by a set of parameters $h$ and we will formulate a joint probability density $p(x, h)$.

A standard result in statistical mechanics is that the equilibrium distribution of the $(x, V)$ system is given by replacing the potential energy $U$ with the enthalpy $U + PV$ in the Boltzmann distribution in Eq.~\eqref{eq:boltzmann_distribution}, which gives:
\begin{equation}
p(x, V) = \frac{1}{Z_P} \exp\Big[-\beta(U(x, V) + P V)\Big],
\label{eq:isotropic_target}
\end{equation}
where $Z_P = \int \mathrm dx \,\mathrm dV \exp[-\beta (U(x, V) + PV)] = \int \mathrm dV \exp[-\beta PV] Z_V$ is the partition function of the isobaric-isothermal ensemble. We  note that the potential energy generally depends on the box shape, hence we include an explicit dependence on $V$. Analogously to the $NVT$ case, the partition function $Z_P$ is related to the \textit{Gibbs free energy} $G$ via
\begin{equation}
G = -\beta^{-1}\nlog{Z_P}.
\end{equation}
The above distribution is directly applicable to simulation boxes whose only shape parameter is $V$, for example a simulation box with isotropic scaling. However, it is not directly applicable to simulation boxes with more than one shape parameter without further assumptions of how the additional shape parameters are distributed.

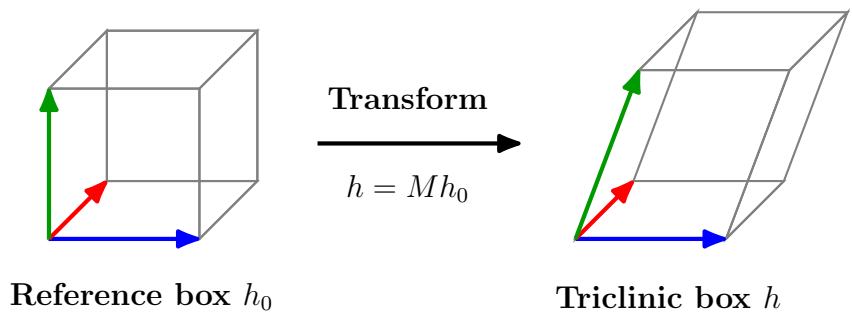
\begin{figure}
    \centering
    \begin{tikzpicture}
    \pgfmathsetmacro\size{2}
    \pgfmathsetmacro\arrowsize{4}
    \pgfmathsetmacro\offset{7}
    \pgfmathsetmacro\height{\size+0.4}
    
    % Reference box
    \coordinate (C0) at (0,0,0);
    \coordinate (C1) at (0,\size,0);
    \coordinate (C2) at (0,\size,\size);
    \coordinate (C3) at (0,0,\size);
    \coordinate (C4) at (\size,0,0);
    \coordinate (C5) at (\size,\size,0);
    \coordinate (C6) at (\size,\size,\size);
    \coordinate (C7) at (\size,0,\size);

    \draw[gray, thick] (C0) -- (C3) -- (C7) -- (C4) -- cycle;
    \draw[gray, thick] (C0) -- (C1) -- (C5) -- (C4) -- cycle;
    \draw[gray, thick] (C0) -- (C1) -- (C2) -- (C3) -- cycle;
    \draw[gray, thick] (C4) -- (C5) -- (C6) -- (C7) -- cycle;
    \draw[gray, thick] (C3) -- (C2) -- (C6) -- (C7) -- cycle;
    \draw[gray, thick] (C1) -- (C2) -- (C6) -- (C5) -- cycle;
    
    \draw [-{Latex[round, blue, length=\arrowsize{}mm]}, blue, ultra thick] (C3) -- (C7);
    \draw [-{Latex[round, red, length=\arrowsize{}mm]}, red, ultra thick] (C3) -- (C0);
    \draw [-{Latex[round, black!40!green, length=\arrowsize{}mm]}, black!40!green, ultra thick] (C3) -- (C2);

    % Triclinic box
    \coordinate (T0) at (\offset,0,0);
    \coordinate (T1) at (\offset+1,\size*1.2,\height-\size);
    \coordinate (T2) at (\offset+1,\size*1.2,\height);
    \coordinate (T3) at (\offset,0,\size);
    \coordinate (T4) at (\offset+\size,0,0);
    \coordinate (T5) at (\offset+1+\size,\size*1.2,\height-\size);
    \coordinate (T6) at (\offset+1+\size,\size*1.2,\height);
    \coordinate (T7) at (\offset+\size,0.,\size);

    \draw[gray, thick] (T0) -- (T3) -- (T7) -- (T4) -- cycle;
    \draw[gray, thick] (T2) -- (T3) -- (T7) -- (T6) -- cycle;
    \draw[gray, thick] (T6) -- (T7) -- (T4) -- (T5) -- cycle;
    \draw[gray, thick] (T2) -- (T6) -- (T5) -- (T1) -- cycle;
    \draw[gray, thick] (T3) -- (T0) -- (T1) -- (T2) -- cycle;

    \draw [-{Latex[round, blue, length=\arrowsize{}mm]}, blue, ultra thick] (T3) -- (T7);
    \draw [-{Latex[round, red, length=\arrowsize{}mm]}, red, ultra thick] (T3) -- (T0);
    \draw [-{Latex[round, black!40!green, length=\arrowsize{}mm]}, black!40!green, ultra thick] (T3) -- (T2);
 
    % Add labels
    \node[font=\bf] at (2,0,4) {Reference box $h_0$};
    \node[font=\bf] at (\offset+2, 0, 4) {Triclinic box $h$};
 
    % Draw arrow
    \draw[-{Latex[round, black, length=\arrowsize{}mm]}, ultra thick] (2.8, 0.5) -- (5.5,0.5);
    \node[font=\bf] at (4, 1.1) {Transform};
    \node[font=\bf] at (4, -0.1) {$h = M h_0$};

\end{tikzpicture}
    \caption{\label{fig:box}Illustration of the box transformation. The cubic reference box $h_0$ \update{(left)} is transformed into a general triclinic box $h$ \update{(right)} by multiplication with an upper triangular matrix $M$ with positive diagonal elements.}
\end{figure}

In this work, we are interested in the case of a triclinic simulation box in $D$ dimensions.
The full flexibility to adjust the box shape is, for example, important when studying solids, to ensure that they can relax to their equilibrium shape~\cite{Parrinello1981, Vega2008}.
To describe the triclinic box shape, we let $h$ be a $D\times D$ matrix with positive determinant that transforms the hypercube $[0, 1]^D$ into the simulation box. For $D=3$, such a simulation box has $9$ shape parameters controlling the following $9$ degrees of freedom: scaling each of the 3 side lengths, varying each of the 3 internal angles, and 3 rigid rotations (see Fig.~\ref{fig:box}). Assuming that all simulation boxes with the same volume $V$ are a priori equally likely, Martyna et al.~\cite{Martyna1994constant} derive the following distribution for the $(x, h)$ system (equation 2.15 in their paper):
\begin{equation}
p(x, h) = \frac{1}{Z_P} \exp\Big[-\beta (U(x, h) + PV) + (1-D)\nlog{V}\Big],
\end{equation}
where $V = \det{h}$ is the volume of the simulation box.

Since the systems we consider are globally rotationally symmetric, it is desirable to eliminate the $D(D-1) / 2$ rigid rotations from the shape parameters $h$. This can be done by restricting $h$ to be upper-triangular ($h_{ij} = 0$ for $i>j$) and requiring its diagonal elements $h_{ii}$ to be positive. In this case, Martyna et al.~\cite{Martyna1994constant} derive the following distribution (equation 2.35 in their paper):
\begin{equation}
p(x, h) = \frac{1}{Z_P} \exp\Bigg[-\beta (U(x, h) + P V) + (1-D)\nlog{V} + \sum_{i=1}^D(i-1)\nlog{h_{ii}}\Bigg].
\end{equation}
where $V = \prod_{i=1}^D h_{ii}$ is the volume of the simulation box.

A practical issue with the above distribution is that the range of atom positions $x = (x_1, \ldots, x_N)$ is not fixed, but varies with the shape parameters $h$. To avoid this complication, we define a fixed reference box with parameters $h_0$ (where we choose $h_0$ to be a diagonal $D\times D$ matrix with positive diagonal elements) and work with reference coordinates $s = (s_1, \ldots, s_N)$ that are defined with respect to the reference box. Concretely, we make the following variable changes:
\begin{enumerate}
    \item $h = M h_0$, where $M$ is a $D\times D$ upper-triangular matrix with positive diagonal elements, and
    \item $x_n = M s_n$ for $n = 1, \ldots, N$.
\end{enumerate}
In addition, we log-transform the diagonal elements of $M$, so that we work with unconstrained shape parameters and that we ensure the diagonal elements $M_{ii}$ are always positive. That is, we define a $D\times D$ upper-triangular matrix $m$ with unconstrained upper-triangular elements, and make the variable change:
\begin{enumerate}
    \setcounter{enumi}{2}
    \item $M_{ii} = \exp(m_{ii})$ for $i = 1, \ldots, D$, and
    \item $M_{ij} = m_{ij}$ for $i\neq j$.
\end{enumerate}
The above transformations change the variables of interest from $(x, h)$ to $(s, m)$. As shown in \ifincludeappendix Appendix \ref{sec:change_of_variables}, \else the Appendix, \fi this change of variables leads to the following probability density for $(s, m)$, which is the target distribution we use in our experiments:
\begin{align}
p(s, m) = \frac{1}{Z_P} \exp\Bigg[&-\beta (U(x, h) + P V) + (N + 2 - D)\nlog{V} \notag 
\\&- (N+1-D) \nlog{V_0} + \sum_{i=1}^D(i-1)\nlog{h_{ii}}\Bigg],
\label{eq:final_npt_target}
\end{align}
where $V = \prod_{i=1}^D \exp(m_{ii})h_{0,i}$ is the volume of the simulation box and $V_0 = \prod_{i=1}^D h_{0,i}$ is the volume of the reference box.
The term $(N+1-D) \nlog{V_0}$ is a constant with respect to $(s, m)$, so it can be absorbed into the normalizing constant.

\subsection{Model architecture\label{sec:npt_flow}}

\begin{figure}
\centering
\includegraphics[width=\textwidth]{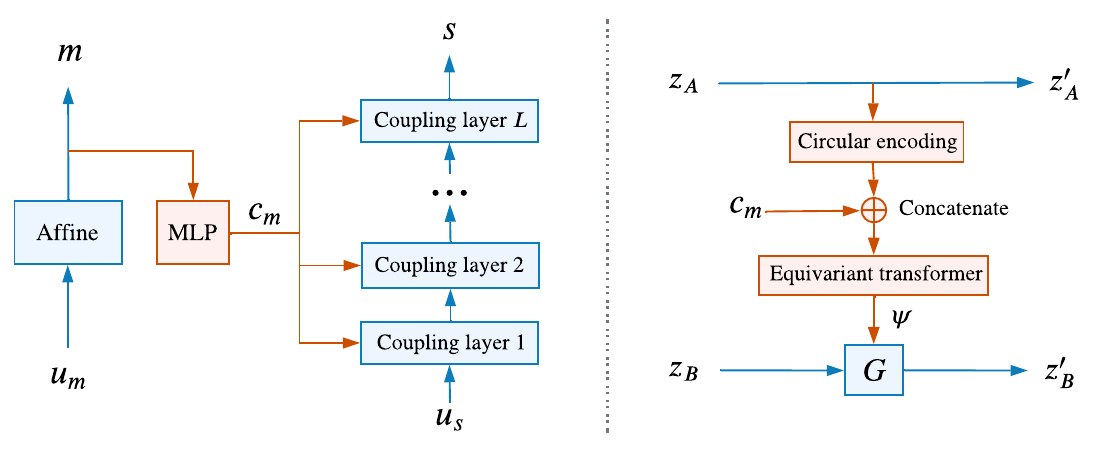}
\caption{\label{fig:npt_flow} \textbf{Left}: High-level architecture of the composite flow model. Blue indicates invertible pathways, red indicates conditional (non-invertible) pathways. The flow over the configuration $s$ also includes other invertible layers (circular shifts) interleaved with the coupling layers, not shown here for simplicity (these layers are not affected by $c_m$). \textbf{Right}: Architecture of a single coupling layer, showing how it is conditioned on the encoded shape $c_m$.}
\end{figure}

In this section, we describe the architecture of a model $q_\theta(s, m)$ that is suitable for approximating the target in Eq.~\eqref{eq:final_npt_target}. Our primary design choice is to decompose the model as:
\begin{equation}
    q_\theta(s, m) = q_{\update{\theta_1}}(m) q_{\update{\theta_2}}(s \g m),
\end{equation}
where $q_{\update{\theta_1}}(m)$ is a flow model over the shape parameters, and $q_{\update{\theta_2}}(s \g m)$ is a \textit{shape-conditional} flow model over the configuration. \update{The trainable parameters of the full model are $\theta = (\theta_1, \theta_2)$, where $\theta_1$ are the trainable parameters of $q_{\theta_1}(m)$ and $\theta_2$ are the trainable parameters of $q_{\theta_2}(s \g m)$.} In the following paragraphs, we describe each model in detail. The overall model architecture is illustrated in Fig.~\ref{fig:npt_flow}.

\subsubsection{Flow over shape parameters}

As previously described in Section~\ref{sec:isothermal_isobaric_ensemble}, $m$ consists of \mbox{$D(D+1)/2$} unconstrained shape parameters. For concreteness, we will focus on $D=3$ (which is the setting of our experiments), in which case $m\in\mathbb{R}^6$. Since $m$ is unconstrained, any standard flow can be used to model its distribution. In our experiments, we use a flow model with base distribution $q_m(u_m)$ and transformation $f_m(u_m)$ defined as follows.
\begin{itemize}
    \item The base distribution is a standard $6$-dimensional Gaussian with zero mean and unit covariance: $q_m(u_m) = \gauss(u_m; 0, I)$.
    \item The flow transformation is affine: $f_m(u_m) = Wu_m + b$, where $W$ is a $6\times 6$ upper-triangular matrix with positive diagonal elements and $b$ is a $6$-dimensional vector. We make the diagonal elements of $W$ positive by parameterizing them as the softplus of unconstrained parameters.
\end{itemize}
The trainable parameters $\update{\theta_1}$ of the \update{above} flow are: the bias $b$, the (pre-softplus) diagonal of $W$, and the above-diagonal elements of $W$. 

This flow parameterizes a $6$-dimensional Gaussian distribution with mean $b$ and covariance $WW^\top$, that is, $q_{\update{\theta_1}}(m) = \gauss(m; b, WW^\top)$. This Gaussian parameterization is complete, in the sense that any Gaussian distribution can be written this way. We found that this Gaussian parameterization is sufficient for our systems of interest (cubic and hexagonal ice): in Section~\ref{sec:shape_parameters_distribution}, we show empirically that the true distribution of shape parameters is statistically indistinguishable from a Gaussian. However, it is straightforward to use more flexible flow models if Gaussianity does not hold.

\subsubsection{Shape-conditional flow over configuration}

The atom configuration $s$ consists of the $ND$ atom coordinates with respect to the orthorhombic reference box $h_0$. Conditional on the shape parameters, the atom configuration is distributed as $p(s \g m) \propto \exp[-\beta U(x, h)]$, which is the Boltzmann distribution of a canonical ensemble confined in the reference box. Therefore, the configuration model $q_{\update{\theta_2}}(s \g m)$ is tasked with approximating a family of canonical ensembles, indexed by the shape parameters $m$.

A general strategy for designing a conditional flow is to start from an unconditional one, $s = f_s(u_s)$, and inject the conditioning information $m$ as an additional input, such that $s = f_s(u_s; m)$. The transformation $f_s(u_s; m)$ must be invertible with respect to $u_s$, but not with respect to $m$. This is the approach we follow in this work.

We build upon the work in Ref.~\cite{Wirnsberger2022normalizing}, which proposed a flow model for atomic solids at fixed volume\footnote{Code was made available at \url{https://github.com/deepmind/flows_for_atomic_solids}}. This flow model consists of a number of invertible \textit{coupling layers}, each of which transforms part of its input (see Fig.~\ref{fig:npt_flow}). Concretely, let $z = (z_{n,i})^{n=1:N}_{i=1:D}$ denote the input to a coupling layer and $z' = (z'_{n,i})^{n=1:N}_{i=1:D}$ denote the output, where $n$ ranges over particles and $i$ ranges over spatial dimensions. The coupling layer works as follows.
\begin{itemize}
    \item First, the input is split in two parts along the $i$ index, $z_A$ and $z_B$. The first part stays fixed, whereas the second gets transformed by an invertible transformation $G(z_B; \psi)$ parameterized by $\psi$. For simplicity, let's assume that the first $d$ spatial dimensions stay fixed, so $z_A = (z_{n,i})^{n=1:N}_{i=1:d}$.
    \item The parameters $\psi$ are computed as a function of $z_A$, $\psi = C(z_A)$, where $C$ is referred to as the \textit{conditioner}. The first step of $C$ is to apply a circular encoding to each particle $(z_{n,i})_{i=1:d}$ (see Appendix A of \cite{Wirnsberger2022normalizing} for details). The second step is to apply a permutation-equivariant transformer architecture \cite{Vaswani2017} to the circular encodings, which outputs the parameters $\psi$.
\end{itemize}
This flow architecture results in a distribution \update{over the configuration $s$} that is invariant to particle permutations.

To make the above architecture shape-conditional, we first encode the shape parameters $m$ into a code $c_m$ using a multi-layer perceptron, as shown in Fig.~\ref{fig:npt_flow}. Then, we make the code $c_m$ an additional input to every conditioner in the flow such that, for every coupling layer, $\psi = C(z_A, c_m)$. We achieve this by concatenating $c_m$ with the circular encoding of each particle $(z_{n,i})_{i=1:d}$ prior to feeding the latter to the permutation-equivariant transformer. This modification preserves the permutation equivariance, and results in a conditional distribution $q_{\update{\theta_2}}(s\g m)$ that is invariant to particle permutations for every $m$.

\update{The trainable parameters $\theta_2$ of this model are: (a) the weights and biases of the multi-layer perceptron that encodes $m$ into $c_m$; (b) the trainable parameters of the permutation-equivariant transformers that parameterize the conditioners $C$ (there is one such transformer for each coupling layer, each with its own trainable parameters); and (c) the trainable circular shifts that are interleaved between coupling layers (these were omitted from this section for simplicity as they are not affected by conditioning on $c_m$, but their details can be found in Appendix A of Ref.~\cite{Wirnsberger2022normalizing}).}

\section{Results\label{sec:results}}
We assess the performance of the $NPT$-flow model proposed in Section~\ref{sec:method} by comparing selected observables and Gibbs free energy estimates to results obtained with \update{MD} simulations. We benchmark our trained models on ice, using a coarse-grained potential called \update{monatomic water}~\cite{Molinero2009}, which treats molecules as point particle interacting via two-body and three-body terms. More specifically, we trained separate models for varying numbers of particles (64, 216 and 512) in the hexagonal (Ih) and cubic (Ic) phases of ice, totalling six experiments with a minimum of ten random seeds each. The goal of the flow model is to draw samples at ambient pressure (1~atm) and a temperature of 200~K. All models were initialized with a perfect lattice at a density of about $1.004~\text{g/cm}^3$, as in Ref.~\cite{Wirnsberger2022normalizing}, which leads to a pressure of approximately $1550~\text{atm}$ in an $NVT$ simulation.

We ran MD in the $NPT$-ensemble using the simulation package LAMMPS~\cite{Plimpton1995}, and employed a Langevin thermostat to control the temperature and a fully flexible barostat to control the pressure~\cite{Shinoda2004, Martyna1994constant, Parrinello1981}\footnote{This can, for example, be done using the combination of the two LAMMPS commands \texttt{fix langevin} and \texttt{fix nph}~\cite{Freitas2016}.}. All further simulation settings are identical to the ones used in Ref.~\cite{Wirnsberger2022normalizing} and we refer to Appendices C--D in the Supplemental Data of that work for further details.

\subsection{Distributions of shape parameters\label{subsec:box}\label{sec:shape_parameters_distribution}}

\begin{figure}
\centering
\includegraphics[width=\textwidth]{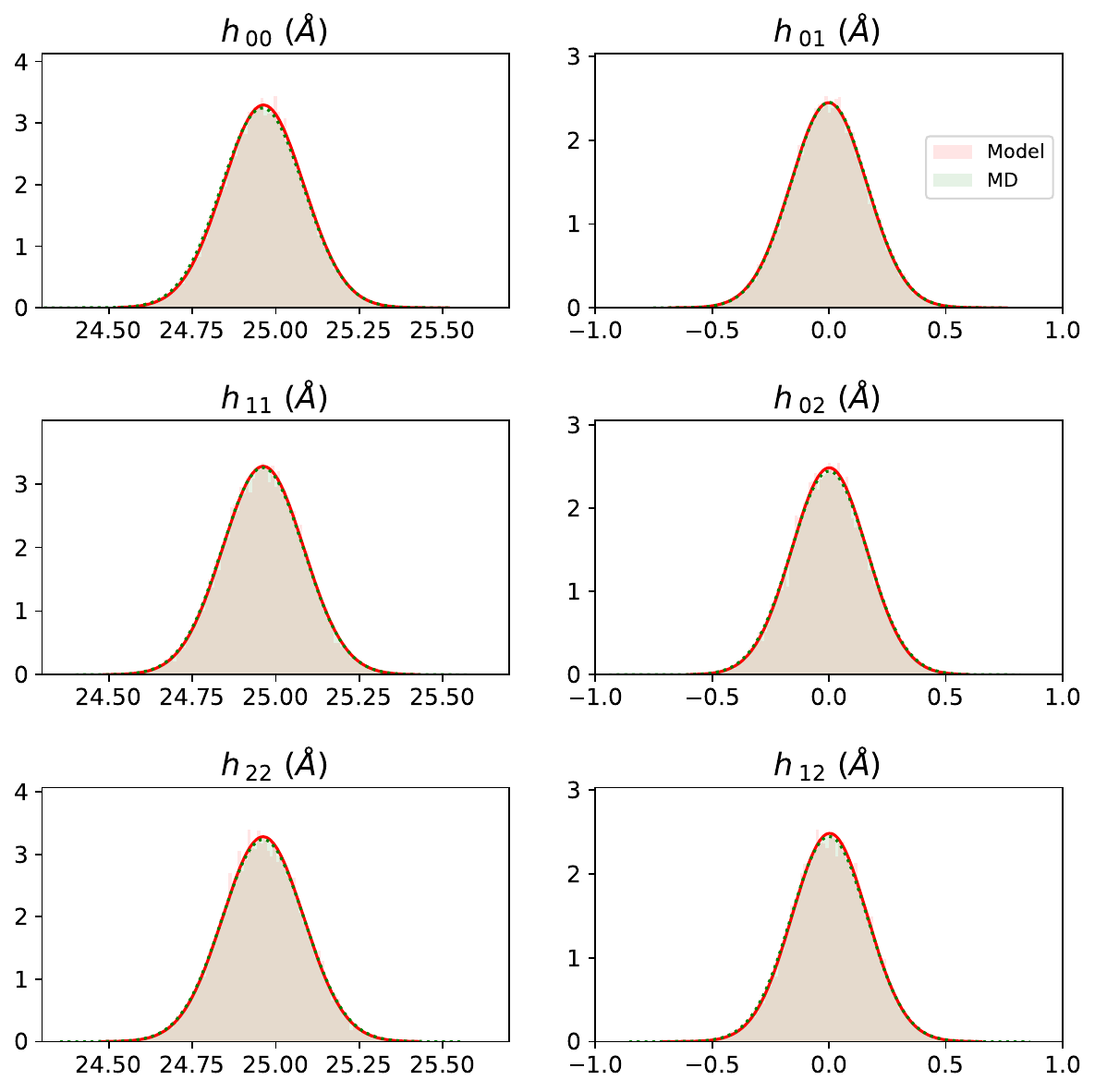}
\caption{\label{fig:npt_box_matrix}Distributions of shape parameters. The panels show the distributions of diagonal (left) and off-diagonal elements of the shape matrix estimated for samples from the trained model (model) and $NPT$ simulations (MD) for the cubic 512-particle system.}
\end{figure}

We first justify our use of a Gaussian-affine flow by verifying that for our systems of interest the true distribution of shape parameters is statistically indistinguishable from a $6$-dimensional Gaussian. To this end, we compare histograms of the six shape parameters sampled from the flow model with their counterparts sampled during MD\@. The results for the 512-particle cubic system are shown in Fig.~\ref{fig:npt_box_matrix} and demonstrate excellent agreement of model and MD\@. Results for hexagonal ice show the same quality of agreement (see \ifincludeappendix Fig.~\ref{fig:npt_box_matrix_hex}  in the \else \fi Appendix) and results for all smaller systems are omitted for the sake of brevity, as training with fewer particles is in general much simpler due to the reduced number of degrees of freedom.

\begin{figure}
\centering
\hspace{2cm}\textbf{Cubic}\hspace{6cm}\textbf{Hexagonal}\vspace{-0.2cm}
\includegraphics[width=0.48\textwidth]{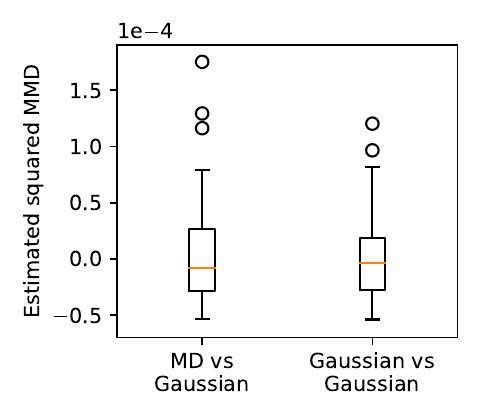}
\includegraphics[width=0.48\textwidth]{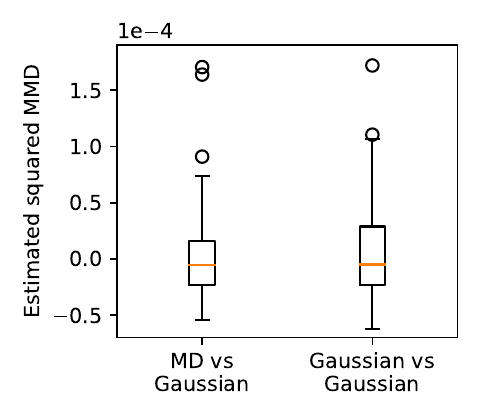}
\caption{\label{fig:mmd}Estimated squared Maximum Mean Discrepancy (using a squared exponential kernel) between the MD distribution of shape parameters and a $6$-dimensional full-covariance Gaussian fit, for cubic (left) and hexagonal (right) ice. The squared MMD was estimated using $10{,}000$ samples, and the experiment was repeated $100$ times (resulting in $100$ MMD estimates). For reference, MMD estimates between the $6$-dimensional Gaussian and itself are also shown. \update{Boxes range from the first to the third quartile of data, with median at the red line; whiskers extend from the corresponding quartile to the farthest datapoint within 1.5 times the box size. All other points are depicted separately as outliers.}}
\end{figure}

To further support our assumption that a Gaussian with full covariance matrix is sufficient to model the shape distribution accurately for ice, we estimated the Maximum Mean Discrepancy (MMD) \cite{Gretton2012mmd} between the MD distribution of shape parameters and a full-covariance Gaussian fit. We observed that MMD estimates are close to zero (zero MMD implies equality of distributions) and that they are distributed similarly to reference MMD estimates between two identical $6$-dimensional Gaussians (see Fig.~\ref{fig:mmd}).

\subsection{Distributions of energy and pressure\label{subsec:energy_pressure}}
\begin{figure}
\centering
\textbf{Cubic}\par
\includegraphics[width=0.48\textwidth]{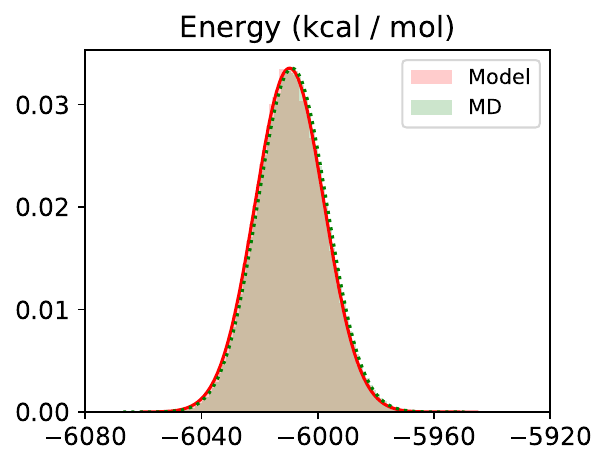}
\includegraphics[width=0.48\textwidth]{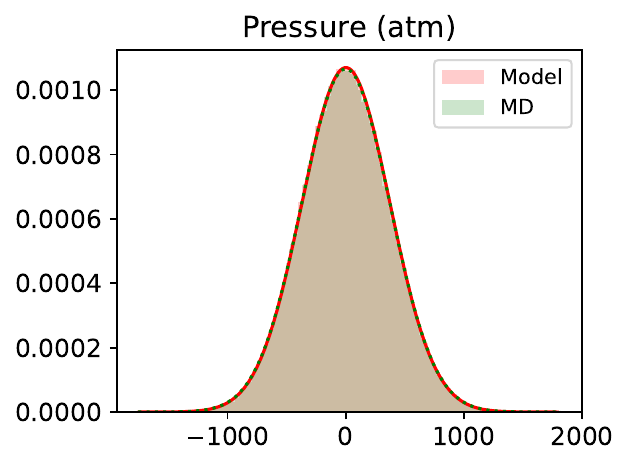}\\
\vspace{0.5cm}
\textbf{Hexagonal}\par
\includegraphics[width=0.48\textwidth]{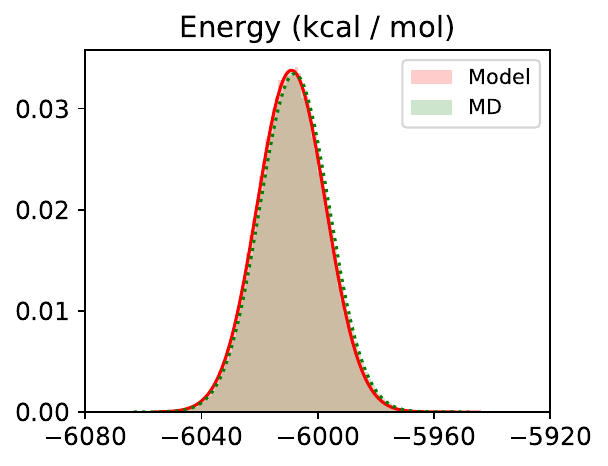}
\includegraphics[width=0.48\textwidth]{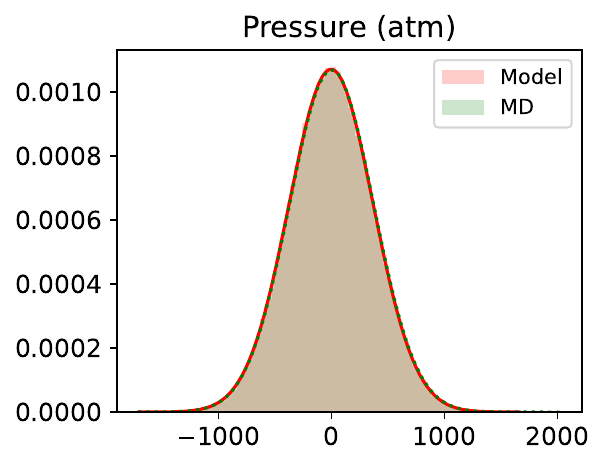}
\caption{\label{fig:energy_pressure}\update{A comparison of the energy histograms (left) and instantaneous pressure histograms (right) for the trained 512-particle cubic (top) and hexagonal (bottom) models and MD\@.}
}
\end{figure}

We next compare the distributions of energies and instantaneous pressure values for both cubic and hexagonal 512-particle systems in Fig.~\ref{fig:energy_pressure} and observe again good overall agreement between model and MD for both quantities. 
\update{Table \ref{tab:observables} shows the average energy and pressure for various system sizes (64, 216 and 512 particles) for both cubic and hexagonal ice, as estimated (a) directly by model samples (biased), (b) by re-weighted model samples (unbiased), and (c) by MD\@. We also show the ESS of importance sampling in each case as a percentage of the actual sample size. The results suggest that the bias due to the model is generally small, and that both observables are accurately estimated. We can further see that importance sampling is able to remove small biases, although at the expense of increasing the error bars, especially for the largest system size (512 particles). Finally, we see that the ESS is reasonably high for 64 and 216 particles, but becomes small for 512 particles, leading to the pronounced increase in estimation uncertainty mentioned previously.}

\begin{table}[ht]
\begin{indented}
\item[]
\caption{\update{Biased (BE) and unbiased importance-sampling (IS) estimates of mean energy and pressure for system sizes of $N=64$, $216$ and $512$ particles, along with estimates of the ESS expressed as a percentage of the actual sample size.
MD estimates were computed with 100 seeds and 10,000 samples each. Model estimates comprise 16 seeds for each phase for 64 particles, 16 and 14 seeds for cubic and hexagonal, respectively, for 216 particles, and 11 seeds for 512 particles, with 512,000 samples per seed in all cases.}\label{tab:observables}
}
\footnotesize\rm
\hspace{-0.65cm}
\begin{tabular}{@{}lrrrrr@{}}
\toprule
&  &  \multicolumn{2}{c}{\textbf{Cubic}}     & \multicolumn{2}{c}{\textbf{Hexagonal}}\\ \midrule
&  &  Energy [kcal/mol]  & Pressure [atm]    &    Energy [kcal/mol] & Pressure [atm] \\ \midrule
\multicolumn{2}{l}{$N=64$} & \multicolumn{2}{c}{ESS = $53.7\%$} &  \multicolumn{2}{c}{ESS = $60.3\%$} \\ 
& BE       & $-750.61 \pm 0.02\hphantom{0}$ & $0.88 \pm 1.10$ & $-750.50 \pm 0.02\hphantom{0}$ & $0.28 \pm 1.98$ \\
& IS &     $-750.456 \pm 0.005$ & $1.15 \pm 0.92$ & $-750.351 \pm 0.005$ & $0.29 \pm 1.98$ \\
& MD       & $-750.451 \pm 0.008$ & $0.85 \pm 0.63$ & $-750.352 \pm 0.009$ & $1.30 \pm 0.73$ \\ \midrule
\multicolumn{2}{l}{$N=216$} & \multicolumn{2}{c}{ESS = $6.8\%$} &  \multicolumn{2}{c}{ESS = $13.6\%$} \\ 
& BE       & $-2535.08 \pm 0.05$ & $1.58 \pm 1.26$ & $-2534.69 \pm 0.04$ & $1.03 \pm 1.46$ \\
& IS       & $-2534.50 \pm 0.11$ & $2.43 \pm 2.67$ & $-2534.20 \pm 0.01$ & $-0.01 \pm 0.80$ \\
& MD       & $-2534.54 \pm 0.02$ & $0.93 \pm 0.39$ & $-2534.22 \pm 0.02$ & $0.80 \pm 0.42$ \\ \midrule
\multicolumn{2}{l}{$N=512$} & \multicolumn{2}{c}{ESS = $0.17\%$} &  \multicolumn{2}{c}{ESS = $0.15\%$} \\ 
& BE       & $-6009.56 \pm 0.17\hphantom{0}$   &  $1.57 \pm 0.91$  & $-6009.07 \pm 0.11\hphantom{0}$    &  $0.93 \pm 1.14$   \\
& IS       & $-6008.84 \pm 0.48\hphantom{0}$   &  $-1.63 \pm 16.1$ & $-6007.6 \pm 1.1\hphantom{00}$      &  $-23.6 \pm 38.8$  \\
& MD       & $-6008.776 \pm 0.027$ &   $0.98 \pm 0.28$ & $-6008.011 \pm 0.024$  &  $0.93 \pm 0.24$   \\
\bottomrule
\end{tabular}
\end{indented}
\end{table}

\subsection{Radial distribution functions\label{subsec:rdf}}
\begin{figure}
\centering
\hspace{1cm}\textbf{Cubic}\hspace{6cm}\textbf{Hexagonal}\par
\includegraphics[width=0.48\textwidth]{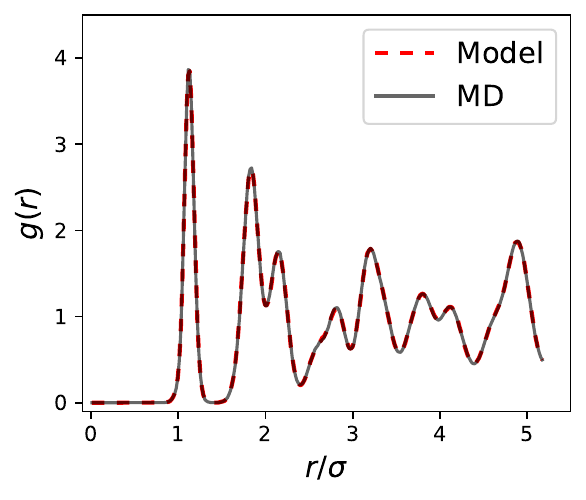}
\includegraphics[width=0.49\textwidth]{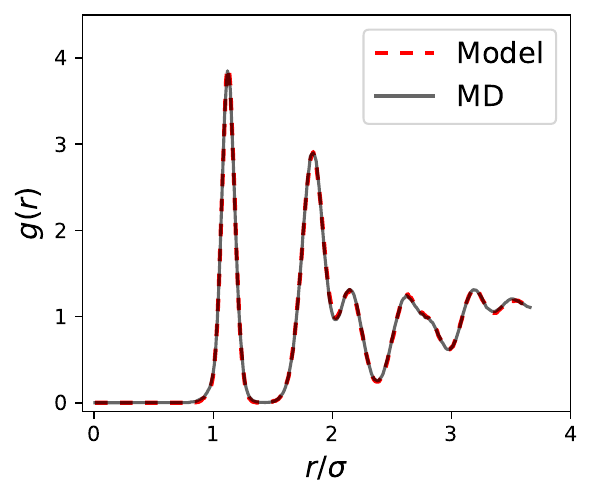}
\caption{\label{fig:rdf} Radial distribution functions for the 512-particle cubic system (left) and hexagonal system (right). Results obtained with the respective models are shown as dashed lines and the MD results are shown as solid lines.
}
\end{figure}

To further assess the quality of the trained model, we computed the \update{radial distribution function} and compare the results for both phases to MD in Fig.~\ref{fig:rdf}. We find the \ourmodel{} results to be almost indistinguishable from MD under this metric, suggesting that the model captures pairwise, structural properties correctly.

\subsection{Gibbs free energies\label{subsec:dG}}
We finally compare Gibbs free energy estimates obtained with our trained models with MD estimates for all three system sizes. For the MD baseline estimates, we first borrow the Helmholtz free energy estimates published in Ref.~\cite{Wirnsberger2022normalizing} and promote these to Gibbs free energy estimates following the method proposed by Cheng and Ceriotti \cite{Cheng2018}. The Gibbs free energy is related to the Helmholtz free energy by
\begin{equation}
    \label{eq:dFtodG}
    G(P, T) = F(h_\text{ref}, T) + P \det(h_\text{ref}) + k_\text{B} T \nlog p (h_\text{ref} \g P, T),
\end{equation}
where $h_\text{ref}$ defines the fixed simulation cell for which we know the Helmholtz free energy estimate, and $\nlog p (h_\text{ref} \g P, T)$ denotes the log-density of the reference box at fixed pressure and temperature. To obtain an estimate of this quantity, we first fit a $6$-dimensional Gaussian to the shape data sampled in an $NPT$-simulation, and then evaluate the log-density of the fit on the reference box analytically.

Table~\ref{tab:fe} contains the Gibbs free energy differences predicted by our models, using the LFEP estimator in Eq.~\eqref{eq:fep}, with the MD baseline estimates computed using Eq.~\eqref{eq:dFtodG}. We find excellent agreement of both methods across all system sizes, verifying that our \ourmodel{} can resolve even the fine free energy difference between cubic and hexagonal ice accurately.

\begin{table}[ht]
\caption{\label{tab:fe}
Gibbs free energy differences between cubic and hexagonal ice, $\mathrm \Delta G = G_\text{cubic} - G_\text{hex}$, at ambient pressure and a temperature of 200~K, reported in units of J/mol.}
\begin{indented}
\item[]
\begin{tabular}{rrr}
\toprule
$N$      &   LFEP         & Baseline   \\ \midrule
64      &   51.76(3)      &  51.64(27)   \\    
216     &   15.61(21)     &  15.56(22)   \\
512     &   6.70(32)      &  6.51(57)    \\
\bottomrule
\end{tabular}
\end{indented}
\end{table}

\section{Discussion\label{sec:discussion}}

To summarize our work, we have proposed a normalizing-flow model, which we call \ourmodel{}, that can be optimized to sample from the isobaric-isothermal ensemble. We have further presented experimental evidence to show that the \ourmodel{} can sample ice for system sizes of up to 512 particles with high accuracy without requiring ground-truth samples to train. All structural and energetic metrics considered in this work agree well with the MD baseline results, even without applying an explicit re-weighting step to unbias estimates. We further demonstrated that the \ourmodel{} can capture the relatively small Gibbs free energy differences between cubic and hexagonal ice accurately when used as a targeted free energy estimator.

We find the agreement of the $NPT$-flow with MD to be of similar quality as the $NVT$-results presented in Ref.~\cite{Wirnsberger2022normalizing}, where the particle-only flow model was used for generating particle coordinates. This suggests that generating the additional 6 shape parameters of the box distribution simultaneously does not complicate the learning task appreciably. Furthermore, the required hardware, training times and quality of results are all comparable with what was reported previously for the canonical ensemble~\cite{Wirnsberger2022normalizing} \update{and are summarized in \ifincludeappendix Appendix~\ref{app:hardware}\else the Appendix\fi}. An interesting extension for future work would be to extend the $NPT$-flow to rigid bodies, such as explicit water, by combining our model with the model proposed by K\"{o}hler et al.~\cite{Kohler2023}.

There are several limitations of our approach, one being that we currently assume a Gaussian-affine box distribution. While this type of model is fully sufficient for the system under consideration, it is conceivable that other problems may require a more flexible description. Making this part more general, however, is a straightforward extension, as we can replace the affine flow with any standard non-affine one from the flow literature~\cite{Papamakarios2021, Kobyzev2021}.

A more fundamental limitation is the restriction to relatively small system sizes. 
\update{In particular, for the hexagonal 512-particle system, we obtain an ESS of only about 0.15\% (expressed as a percentage of the actual sample size) compared to an ESS of around 60\% for 64 particles (see Tab.~\ref{tab:observables}). This can be overcome with drawing enough samples, which is computationally efficient with a flow model, as sampling can be done in parallel. However, the rapid decay in the statistical efficiency of naive importance sampling as dimensionality increases  makes scaling the approach to larger systems challenging. 
One option for pushing the dimensionality further would be to consider more sophisticated estimation methods than naive importance sampling, in particular multi-stage methods capable of working with flow models~\cite{Wirnsberger2020, Matthews2022, Rizzi2023, Arbel2021, Midgley2023}. Another avenue would be to improve the architecture and training methods of flow models further, but this would require further innovations in the field.}

\update{Finally, although not the focus of this paper, it is worth commenting on the broader comparison between flow-based sampling methods and MD\@. Apart from the fact that MD models the temporal evolution of a system, which makes it applicable to a wider class of problems, a core difference between the two approaches is that flow-based methods can produce independent samples that can be generated in parallel, whereas MD is a sequential method based on small integration steps leading to correlated samples. Therefore, flow-based methods are in principle not constrained by the existence of low-energy barriers, as for example discussed by No\'e et al.~\cite{Noe2019}. Moreover, flow-based sampling is easily parallelizable and therefore well-suited to modern distributed hardware. The main downside of flow-based methods is the need for training, which as discussed becomes increasingly challenging for larger systems. Therefore, a promising avenue for flow-based methods is amortization, where a single flow model may be trained to approximate a range of different states, parameterized, for example, by particle size, temperature, pressure or other quantities of interest, so that the cost of training is amortized across many evaluations.}

\ack
\ifanonymized
Removed for double-anonymous review.
\else
We would like to thank our colleagues S{\'e}bastien Racani{\`e}re, Andrew J.~Ballard, Alexander Pritzel, Danilo Jimenez Rezende, Charles Blundell, Th{\'e}ophane Weber, John Jumper, Jonas Adler, Alexander G.D.G. Matthews and Arnaud Doucet for their help and for stimulating discussions.
\fi

\section*{References}
\bibliographystyle{iopart-num}
\bibliography{references}

\ifincludeappendix

\newcommand{\beginsupplement}{%
        \setcounter{table}{0}
        \renewcommand{\tablename}{Supplementary Table}
        \setcounter{figure}{0}
        \renewcommand{\theequation}{S\arabic{equation}}
        \setcounter{equation}{0}
        \renewcommand{\figurename}{Supplementary Figure}
        \renewcommand{\thefigure}{S\arabic{figure}}
        \renewcommand{\theHfigure}{S\arabic{figure}}
        \setcounter{section}{0}
        \renewcommand\thesection{\Alph{section}}
     }
     
%\clearpage
\beginsupplement

\section{Change of variables for upper-triangular triclinic boxes\label{sec:change_of_variables}}

As described in Section~\ref{sec:isothermal_isobaric_ensemble}, the isobaric-isothermal ensemble for upper-triangular (non-rotating) triclinic simulation boxes is given by
\begin{equation}
p(x, h) = \frac{1}{Z_P} \exp\Bigg[-\beta (U(x, h) + P V) + (1-D)\nlog{V} + \sum_{i=1}^D(i-1)\nlog{h_{ii}}\Bigg],
\end{equation}
where $V = \prod_{i=1}^D h_{ii}$. We want to change variables from $(x, h)$ to $(s, m)$ as described by:
\begin{enumerate}
    \item $h = M h_0$ where $M$ is upper-triangular,
    \item $x_n = M s_n$ for $n = 1, \ldots, N$,
    \item $M_{ii} = \exp(m_{ii})$ for $i = 1, \ldots, D$, and
    \item $M_{ij} = m_{ij}$ for $i\neq j$.
\end{enumerate}
We will apply the general rule for variable changes, $p(x) = p(y)\abs{\det\frac{\partial y}{\partial x}}$. We compute the Jacobian determinant of each of the above transformations as follows.
\begin{enumerate}
    \item We break this transformation into a sequence of $D$ transformations, the $i$-th of which transforms the $i$-th row of $M$ into the $i$-th row of $h$. Each of these is a linear transformation with matrix $h_0^\top$ and Jacobian determinant $\det h_0^\top = V_0$, so the total Jacobian determinant is $V_0^D$.
    \item We have $N$ linear transformations, each with matrix $M$ and Jacobian determinant $\det M = \det h / \det h_0 = V/V_0$, so the total Jacobian determinant is $V^N/V_0^N$. 
    \item We have $D$ scalar transformations, the $i$-th of which has derivative $\exp(m_{ii}) = M_{ii}$. Therefore, the total Jacobian determinant is $\prod_{i=1}^D M_{ii} = \det M = V/V_0$.
    \item This is the identity transformation, so its Jacobian determinant is $1$.
\end{enumerate}
Collecting the four Jacobian determinants in the change-of-variables formula, we get
\begin{align}
p(s, m) = p(x, h) \frac{V^{N + 1}}{V_0^{N + 1 - D}} = \frac{1}{Z_P} \exp\Bigg[&-\beta (U(x, h) + P V) + (N + 2 - D)\nlog{V} \notag\\&- (N+1-D) \nlog{V_0} + \sum_{i=1}^D(i-1)\nlog{h_{ii}}\Bigg].
\end{align}

\section{Additional results}
Figure~\ref{fig:npt_box_matrix_hex} compares the marginal distributions of box shape parameters of the hexagonal 512-particle system with MD, and demonstrates the same quality of agreement as we observed for the cubic system in the main text. 

\begin{figure}
\centering
\includegraphics[width=\textwidth]{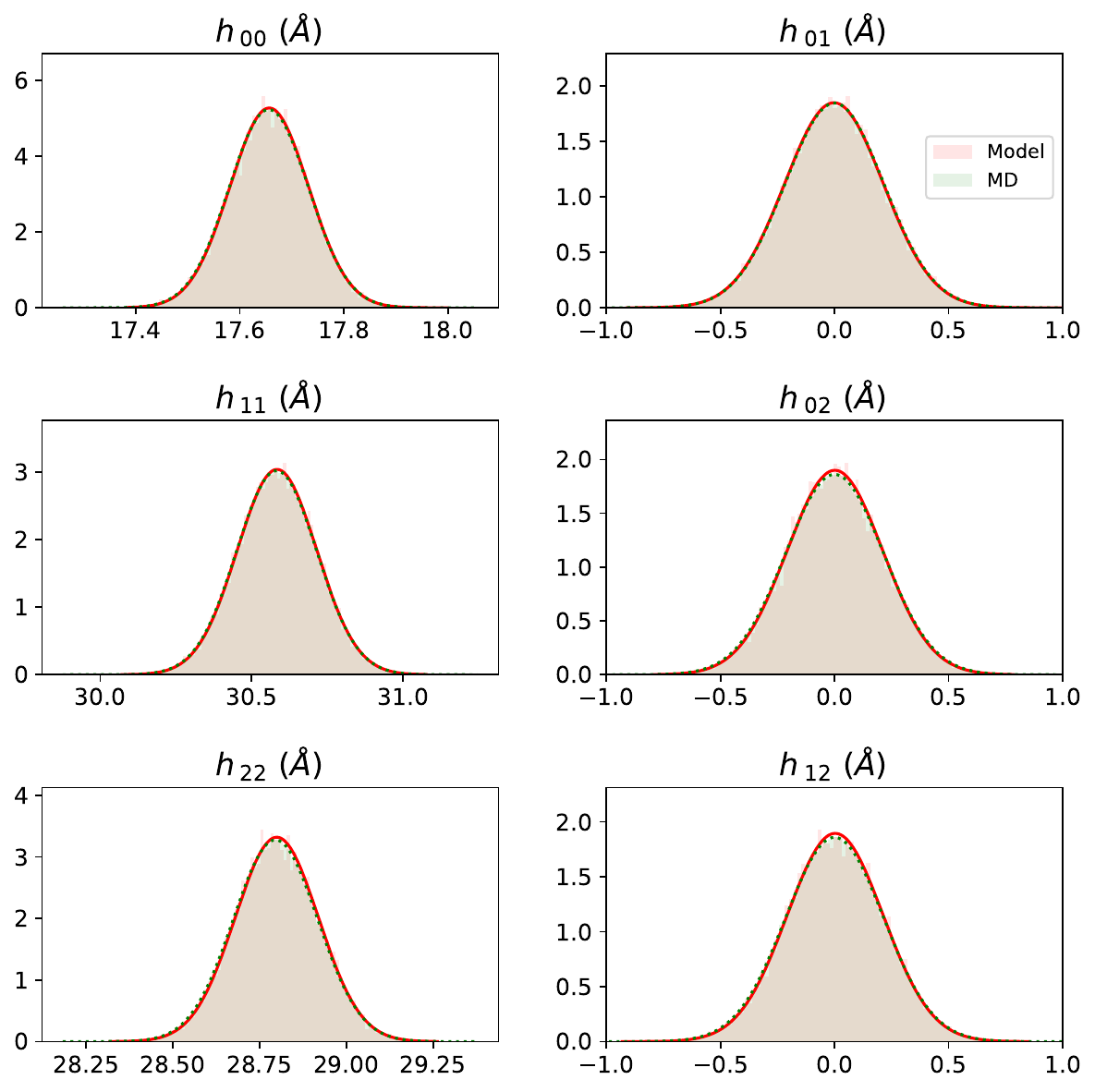}
\caption{\label{fig:npt_box_matrix_hex}The panels show the distributions of diagonal (left) and off-diagonal elements of the shape matrix estimated for samples from the trained model (model) and $NPT$ simulations (MD) for the hexagonal 512-particle system.}
\end{figure}
\update{
\section{\label{app:hardware}Details on hardware and computational cost}
Training until convergence on the small system with 64 particles took about two days on four V100 GPU and generating 512K samples
took about twenty minutes on two V100 GPUs. 
For the medium system with 216 particles, training until convergence took about five days on eight V100 GPUs. Drawing 
512K samples from the model took about one hour on four V100 GPUs. For the largest system with 512 particles, we 
used 16 A100 GPUs for training and sampling. Training took approximately three weeks, and 
drawing 512K samples from the model took about 2 hours.
}

\else
% no appendix
\fi

\end{document}